\documentstyle[emulateapj,onecolfloat,psfig]{article}

\def\veck{{\bf k}}

\long\def\comment#1{}

\def\la{\hbox{ \raise.35ex\rlap{$<$}\lower.6ex\hbox{$\sim$}\ }}
\def\ga{\hbox{ \raise.35ex\rlap{$>$}\lower.6ex\hbox{$\sim$}\ }}

\def\bfk{{\bf k}}
\def\W2{{\cal W}}

\newcommand{\fore}{{\rm f}}
 \newcommand{\wj}{\left(
                          \begin{array}{ccc}
                          l_1  &  l_2  & l_3 \\
                            0  &  0    &  0
                          \end{array}
                          \right)}
\newcommand{\wjm}{\left(
                           \begin{array}{ccc}
         l_1 & l_2  & l_3  \\
         m_1 & m_2  & m_3
                           \end{array}
                   \right)}

\newcommand{\bi}{B_{l_1 l_2 l_3}}
\newcommand{\bilm}{B_{l_1 l_2 l_3}^{m_1 m_2 m_3}}

\newcommand{\deld}{\delta^{\rm D}}
\newcommand{\bn}{\hat{\bf n}}

\newcommand{\bk}{\hat{\bf k}}
\newcommand{\rad}{r}    
\newcommand{\da}{d_A}   
\newcommand{\tableskip}{\tablevspace{3pt}}

\newcommand{\sz}{{\rm SZ}}

\newcommand{\cmb}{{\rm CMB}}

\newcommand{\sky}{{\rm sky}}
\newcommand{\tot}{{\rm tot}}

\newcommand{\Ylm}[1]{Y_{l_#1}^{m_#1}}
\newcommand{\Ylmn}{Y_{l}^{m}}

\newcommand{\almn}{a_{l m}}

\begin{document}
\twocolumn[
\title{
Large-Scale Sunyaev-Zel'dovich Effect: Measuring Statistical Properties with 
Multifrequency Maps}

\author{Asantha Cooray$^1$, Wayne Hu$^2$, and Max Tegmark$^3$}
\affil{
$^1$Department of Astronomy and Astrophysics, University of Chicago,
Chicago IL 60637\\
$^2$Institute for Advanced Study, Princeton, NJ 08540\\
$^3$Department of Physics, University of Pennsylvania, Philadelphia, PA 19104\\
E-mail: asante@hyde.uchicago.edu, whu@ias.edu, max@physics.upenn.edu}
\submitted{Submitted for publication in The Astrophysical Journal}
\begin{abstract}
We study the prospects for extracting detailed statistical properties
of the Sunyaev-Zel'dovich (SZ) effect associated with large 
scale structure using upcoming multifrequency CMB experiments.  
The greatest obstacle
to detecting the large-angle signal is the confusion noise
provided by the primary anisotropies themselves, and to a lesser
degree galactic and extragalactic foregrounds.   We employ multifrequency
subtraction techniques and the latest foregrounds models to determine
the detection threshold for the Boomerang, MAP (several $\mu$K)  
and Planck CMB (sub $\mu$K) experiments.
Calibrating a simplified biased-tracer model of the gas pressure 
off recent hydrodynamic simulations, we estimate the SZ power spectrum,
skewness and bispectrum through analytic scalings and N-body
simulations of the dark matter.  We show that the Planck satellite should
be able to measure the SZ effect with sufficient precision to determine
its power spectrum and higher order correlations, e.g. the skewness
and bispectrum.  Planck should also be able to detect the
cross correlation between the SZ and gravitational lensing effect
in the CMB.  Detection of these effects will help determine the properties 
of the as yet undetected gas, including the manner in which the gas pressure
traces the dark matter.
\end{abstract}


]
\section{Introduction}

It is by now well established that the precision measurements of the cosmic
microwave background expected from upcoming experiments, especially 
MAP and Planck satellite missions, will provide a gold mine of information 
about the early universe and the fundamental cosmological parameters
(e.g., \cite{Junetal96} 1996).  
These experiments can in fact do so much more.  With all-sky maps across
the wide range of uncharted frequencies from $20$GHz-$900$GHz, the secondary science
from these missions will arguably be as interesting as the primary science.

In this paper, we examine the prospects for extracting the large-scale
properties of the hot intergalactic gas from multifrequency observations of the CMB. 
Inverse-Compton scattering of CMB photons by hot gas, known as
the Sunyaev-Zel'dovich (SZ; \cite{SunZel80} 1980) 
effect, leaves a characteristic distortion
in the spectrum of the CMB, which fluctuates in the sky with the gas
density and temperature.  
In the Rayleigh-Jeans (RJ) regime, it produces
a constant decrement and with only low frequency measurements, the
much larger primary anisotropies in the CMB itself obscure the
fluctuations on scales greater than a few arcminutes (e.g., 
\cite{GolSpe99} 1999).  
The upscattering in frequency implies an increment at high frequencies
and a null around $217$GHz.  This behavior provides a potential tool for the separation
of
SZ effect from other temperature anisotropy contributors.

Since both the SZ spectrum and the CMB spectrum are accurately known,
one can expect that foreground removal techniques developed to isolate
the primary anisotropies can be reversed to recover the SZ signal in the presence 
of noise from the primary anisotropies.    Galactic and extragalactic
foregrounds will be more challenging to remove.  Here we use the latest
foreground models from \cite{Tegetal99} (1999) that takes into account the fact
that imperfect correlations in the foregrounds between frequency
channels inhibits our 
ability to remove them. Using foreground information together with the
expected noise properties of individual experiments,
one can determine the  minimal detectable
signal in each experiment and the upper limit
achievable in the absence of
detection.  Experiments with sufficient signal-to-noise can 
extract precision measurements for the power spectrum 
and higher order statistics such as the skewness.  Ultimately, they can
provide detailed maps of the large-angle SZ effect.

To assess the prospects for an actual detection, we must model the
SZ signal itself.  
The SZ effect is now routinely imaged
in massive galaxy clusters (e.g., \cite{Caretal96} 1996; \cite{Jonetal93} 1993),
where the temperature of the scattering medium  can reach as high as
10 keV, producing temperature changes in the CMB of order 1 mK at
RJ wavelengths.  The possibility for detection of massive clusters in
CMB satellite data has already been discussed in several studies
(e.g., \cite{Aghetal96}, \cite{HaeTeg96} 1996, \cite{Poietal98} 1998).
Here, however, we are interested in the SZ effect produced by
large-scale structure in the general intergalactic medium (IGM) where 
the gas is expected to be at $\lesssim 1$keV in mild overdensities, 
leading to CMB contributions in the $\mu$K range. 

It is now widely believed that at least $\sim$ 50\% of the
present day baryons, when compared to the total baryon budget from
big bang nucleosynthesis, are present in gas associated with hot large-scale
structure which has remained undetected given its
temperature and clustering properties
(e.g., \cite{Fuketal98} 1998; \cite{CenOst99} 1999; \cite{Pen99} 1999).  
Recently, \cite{Schetal00} (2000) has provided a tentative detection of
X-ray emission from a large-scale filament in one of the deep
ROSAT PSPC fields; previous attempts 
yielding upper limits are described in
\cite{KulBoh99} (1999) and \cite{BriHen95} (1995). 
These results are consistent with current predictions for the X-ray
surface brightness based on numerical simulations (e.g.,
\cite{Cenetal95} 1995).  
\cite{Pen99} (1999) argued that non-gravitational heating of the
gas to $\sim 1$keV is required to evade bounds from the soft X-ray background.
These results suggest that the X-ray emission from this gas 
may be detectable in the near future with 
wide-field observations with Chandra X-ray
Observatory\footnote{http://asc.harvard.edu} and X-ray Multiple Mirror
Mission\footnote{http://astro.estec.esa.nl/XMM}.

On the theoretical front, hydrodynamic simulations
of the SZ effect continue to improve 
(\cite{daS99} 1999; \cite{Refetal99} 1999; \cite{Seletal00} 2000).  As a consensus from these simulations of basic
properties such as the opacity weighted gas temperature and average
Compton distortion is still lacking, we will base our assessment 
of the detectability of the
large-scale SZ effect on a simple parameterization of the
effect, based on a gas pressure bias model
(\cite{Refetal99} 1999), 
crudely calibrated with the recent hydrodynamic simulations.
We employ perturbation theory, non-linear scaling relations, and N-body simulations for the dark matter to assess the statistical properties of the
signal.
Properly calibrated, these techniques can complement hydrodynamic 
simulations by extending their dynamic range and sampling volume.  
Currently, they should simply be 
taken as order of magnitude estimates of the potential signal.

Throughout this paper, we will take an adiabatic cold dark matter (CDM)
model as our fiducial cosmology.  We assume cosmological
parameters $\Omega_c=0.30$ for the cold dark matter density, 
$\Omega_b=0.05$ for the baryon density, $\Omega_\Lambda=0.65$ for the
cosmological constant, $h=0.65$ for the dimensionless Hubble 
constant and a COBE-normalized scale invariant spectrum of
primordial fluctuations (\cite{BunWhi97} 1997).  

The layout of the paper is as follows.
In \S~\ref{sec:cleaning}, we describe the foreground and primary anisotropy 
removal method and assess their efficacy for upcoming CMB experiments. 
In \S~\ref{sec:sz}, we detail the bias model for the SZ effect
and calculate through perturbation theory, analytic approximations and
numerical simulations, 
the low order statistics of the SZ effect: its power spectrum, skewness 
and bispectrum.
In \S~\ref{sec:sn}, having estimated the noise and the signal, we 
assess the prospects for measuring these low order statistics in 
upcoming experiments. 
We conclude in \S~\ref{sec:discussion} with a discussion of
our results.

\begin{figure*}
\centerline{\psfig{file=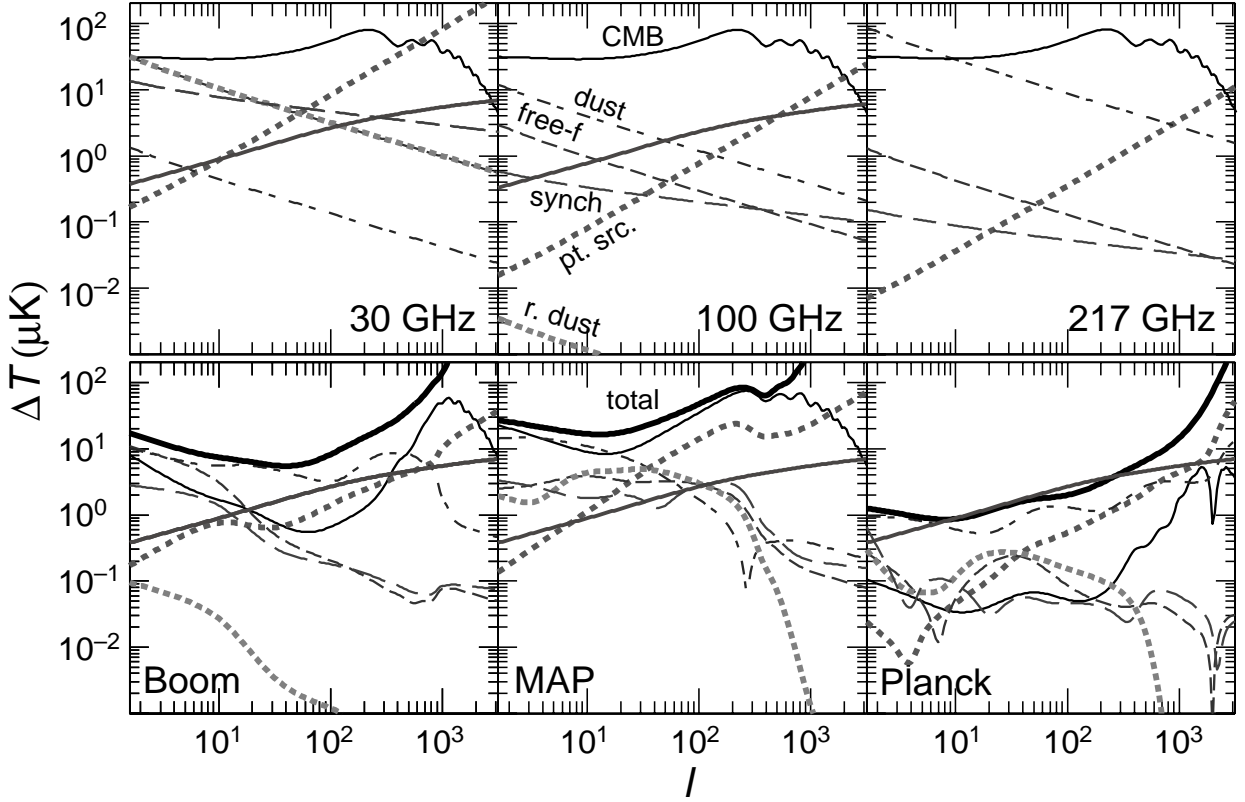,width=6.5in,angle=0}}
\caption{Top: foreground contributions to temperature anisotropies 
$(\Delta T/T)^2 = l(l+1)C_l/2\pi$ from the various foregrounds
(dust, free-free, synchrotron, radio and infrared point sources,
and rotating dust) at three fiducial frequencies as labeled.
The SZ signal (solid, unlabeled) is estimated with the simplified model of \S~3.
Bottom: residual foregrounds after multifrequency subtraction for
Boomerang, MAP and Planck. The total includes detector noise and residual CMB.
}
\label{fig:clean}
\end{figure*}
 
\section{Modeling the CMB and Foreground Noise}
\label{sec:cleaning}

The main obstacle for the detection of the SZ effect from large-scale
structure for angular scales above a few arcminutes is the CMB itself.
Here the primary anisotropies dominate the SZ effect for frequencies
near
and below the peak in the CMB spectrum (see Fig.~\ref{fig:clean}).   
Fortunately, the known
frequency dependence and statistical properties of primary
anisotropies allows 
for extremely
effective subtraction of their contribution (e.g., 
\cite{Hobetal98} 1998; \cite{BouGis99} 1999).  
In particular, primary anisotropies obey Gaussian
statistics
and follow the blackbody spectrum precisely.
 
Perhaps more worrying are the galactic and extragalactic foregrounds,
some of which are expected to to be at least comparable to the SZ
signal in
each frequency band.  These foregrounds typically have spatial and/or
temporal
variations in their frequency dependence leading to imperfect
correlations 
between
their contributions in different frequency bands.   We attempt here to
provide
as realistic an estimate as possible of the prospects for CMB and
foreground
removal, given our incomplete understanding of
the foregrounds.

\begin{table}[tb]\footnotesize
\caption{\label{tab:specs}}
\begin{center}
{\sc CMB Experimental Specifications}
\begin{tabular}{rcccc}
\tableskip\hline\hline\tableskip
Experiment & $\nu$ & FWHM & $10^6 \Delta T/T$ &  \\
\tableskip\hline\tableskip
Boomerang
& 90 & 20 & 7.4 \\
& 150 & 12 & 5.7 \\
& 240 & 12 & 10 \\
& 400 & 12 & 80 \\
\tableskip\hline\tableskip
MAP
& 22 & 56 & 4.1  \\
& 30 & 41 & 5.7  \\
& 40 & 28 & 8.2  \\
& 60 & 21 & 11.0 \\
& 90 & 13 & 18.3 \\
\tableskip\hline\tableskip
Planck
& 30  & 33 & 1.6 \\
& 44  & 23 & 2.4 \\
& 70  & 14 & 3.6 \\
& 100 & 10 & 4.3 \\
& 100 & 10.7 & 1.7 \\
& 143 & 8.0 & 2.0  \\
& 217 & 5.5 & 4.3  \\
& 353 & 5.0 & 14.4 \\
& 545 & 5.0 & 147  \\
& 857 & 5.0 & 6670 \\
\tableskip\hline
\end{tabular}
\end{center}
NOTES.---%
Specifications used for 
Boomerang, MAP and Planck.
Full width at half maxima (FWHM) of the beams are in arcminutes and
should be converted to radians for the noise formula. 
Boomerang covers a fraction $\sim$ 2.6\% of the sky, while we assume
a usable fraction of 65\% for MAP and Planck. In \S~\ref{sec:sn}, in
order to calculate the maximum signal-to-noise, we define a {\it perfect}
experiment as one with no instrumental noise and full sky coverage.
\end{table}

\subsection{Foreground Model and Removal}
\label{sec:foregmodel}

We use the ``MID'' foreground model of 
\cite{Tegetal99} (1999) and adapt the subtraction techniques found
there for the purpose of extracting the SZ signal.  
The assumed level of the foreground signal in the power spectrum
for three fiducial frequencies is shown in Fig.~\ref{fig:clean}.

The foreground model is defined in terms of the covariance between
the multipole moments at different frequency 
bands\footnote{A potential caveat for this type of modeling is that it 
assumes the foregrounds are statistically
isotropic whereas we know that the presence of the Galaxy violates
this assumption at least for the low order multipoles.   We assume that
$1-f_\sky \sim 0.35$ of the sky is lost to this assumption even with
an all-sky experiment. }
\begin{equation}
\left< a_{l' m'}^{\fore *}(\nu') a_{l m}^{\fore} (\nu) \right> = 
C_l^{\fore}(\nu',\nu)
               \delta_{l l'} \delta_{m m'}\,,
\end{equation}
in thermodynamic temperature units as set by the CMB blackbody.  
In this section, we will speak of the primary
anisotropies and detector noise simply as other foregrounds with very
special 
properties:
\begin{eqnarray}
C_l^{\rm CMB}(\nu',\nu)&=& C_l\,, \nonumber\\
C_l^{\rm noise}(\nu',\nu)&=& 8\ln 2 \theta(\nu)^2 e^{\theta^2(\nu) l(l+1)}
	\left({\Delta T\over T}\right)^2\Big|_{\rm noise}
\delta_{\nu,\nu'}\,. \nonumber \\
\label{eqn:clnoise}
\end{eqnarray}
The FWHM$=\sqrt{8\ln 2} \theta$ and noise specifications 
of the Boomerang, MAP and Planck frequency channels 
are given in Tab.~1.  True foregrounds generally fall in 
between these extremes of perfect and no frequency correlation.

The difference between extracting the SZ signal and the primary signal
is 
simply
that one performs the subtraction referenced to the
SZ frequency dependence
\begin{equation}
s(\nu) = 2 - {x \over 2}\coth {x \over 2}\,,
\end{equation}
where $x = h\nu/kT_{\rm cmb} \approx \nu/56.8$GHz.   Note that
in the RJ limit 
$s(\nu) \rightarrow 1$ such that
\begin{equation}
C_l^{\rm SZ}(\nu,\nu') = s(\nu)s(\nu') C_l^{\rm SZ}
\end{equation}
where $C_l^{\rm SZ}$ is the SZ power spectrum in the RJ
limit.
 
Consider an arbitrary linear combination of the channels,
\begin{equation}
b = \sum_{\nu_{i}} {1 \over s(\nu_i)} w(\nu_i) a(\nu_i)\,.
\end{equation}
Since the subtraction is done multipole by multipole, we have
temporarily
suppressed the multipole index.
The covariance of this quantity is
\begin{equation}
\left< b^2 \right> = C^{\sz}[\sum_{\nu_i} w(\nu_i)]^2
        + \sum_{\nu_i,\nu_j} w(\nu_i) w(\nu_j)
        \sum_{\fore} {C^\fore(\nu_i,\nu_j) \over
s(\nu_i)s(\nu_j)} 
\,.
\end{equation}
Minimizing the variance contributed by the foregrounds subject to the
constraint that the SZ estimation be unbiased, we obtain
\begin{equation}
\sum_{\nu_i} w(\nu_j) \sum_{\fore} {C^\fore(\nu_i,\nu_j) \over 
s(\nu_i)s(\nu_j)} =
{\rm const.}\,
\end{equation}
Defining the scaled foreground covariance matrix as
\begin{eqnarray}
\tilde C(\nu_i,\nu_j) &=&
\sum_{\fore} {C^\fore(\nu_i,\nu_j) \over s(\nu_i)s(\nu_j)}, 
\nonumber\\
&\equiv& \sum_{\fore} \tilde C^\fore(\nu_i,\nu_j) \,,
\end{eqnarray}
we solve for the weights that minimize the noise variance
\begin{equation}
{\bf w} \propto {\tilde {\bf C}^{-1} {\bf e}} \, ,
\label{eqn:weights}
\end{equation}
where ${\bf e}$ is the vector of all ones $e(\nu_i)=1$.  Finally we
normalize the sum of the weights to unity $\sum w(\nu_i)=1$ to obtain
an unbiased estimator.
Our approach is same as minimizing the foreground variance subject to
the constraint that the recovered multipole is an unbiased estimate of
the true SZ signal. As each channel is rescaled such that SZ signal
corresponds to the RJ level, the weights sum to unity.

The total residual noise variance in the map from the foregrounds per
multipole 
is then
\begin{equation}
N_l = {\bf w}_l^t {\tilde {\bf C}_l} {\bf w}_l\,,
\label{eqn:nl}
\end{equation}
and from each foreground component
\begin{equation}
N_l^\fore = {\bf w}_l^t {\tilde {\bf C}_l^\fore} {\bf w}_l\,.
\label{eqn:residualcomponent}
\end{equation}
Note that the residual noise in the map is independent of assumptions
about the
SZ signal including whether it is Gaussian or not.   However if the
foregrounds
themselves are non-Gaussian, then this technique only minimizes the
variance
and may not be optimal for recovery of non-Gaussian features in the SZ
map 
itself. \cite{Bouetal95} (1995) have shown that similar techniques are quite
effective 
even
when confronted with non-Gaussian foregrounds.  
This is a potential caveat especially for cases in which 
the residual noise is not dominated by the primary anisotropies 
or detector noise.  We shall discuss methods to alleviate this
concern in the next section.

\begin{figure}[t]
\centerline{\psfig{file=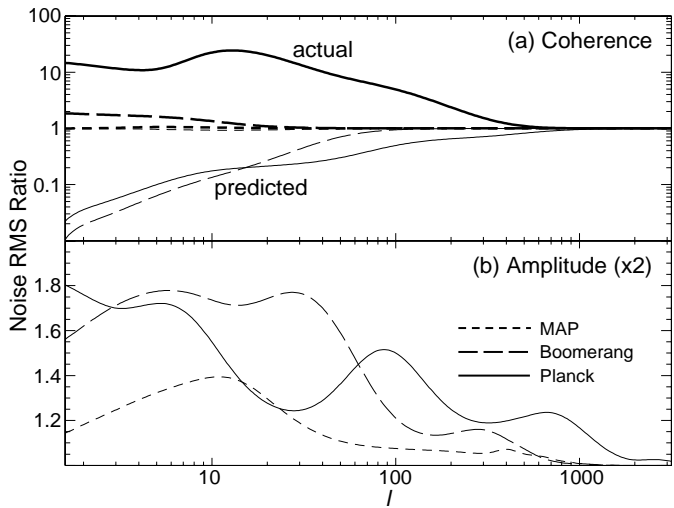,width=3.5in,angle=0}}
\caption{Dependence of the residual noise rms on foreground
assumptions expressed as a ratio to the fiducial model of Fig.~\protect\ref{fig:clean}.  (a) Falsely assuming the foregrounds have perfect
frequency coherence not only underpredicts the residual noise by a substantial
factor but also leads to substantially more actual residual noise.
(b) Multiplying the foreground amplitudes by 2 (power by
4) produces less than a factor of 2 increase in the residual noise.
}
\label{fig:simul}
\end{figure}
\subsection{Detection Threshold}

The residual noise sets the detection threshold for the
SZ effect for a given experiment.  
In Fig.~\ref{fig:clean}, 
we show the rms of the residual noise after
foreground subtraction for the Boomerang, MAP and Planck
experiments assuming the ``MID'' foreground model from
\cite{Tegetal99} (1999).  With the Boomerang and Planck channels,
elimination of the primary anisotropies is excellent up to the beam
scale where detector noise dominates.  As expected, the MAP channels,
which are all on the RJ side of the spectrum, do not
allow good elimination of the primary anisotropies.

It is important not to assume that the foregrounds are 
perfectly correlated in frequency, which is the usual 
assumption in the literature (\cite{Hobetal98} 1998; 
\cite{BouGis99} 1999).  There are two types of errors
incurred by doing so.  The first is that one underpredicts
the amount of residual noise in the SZ map (see Fig.~\ref{fig:simul}).
The second is that if one calculates the optimal weights in
equation~(\ref{eqn:weights}) based on this assumption the actual
residual noise increases.  For Planck it can actually increase the noise
beyond the level in which it appears in the $100$GHz maps with no
foreground subtraction at all.  The reason is that the cleaning 
algorithm then erroneously uses the contaminated high and low
frequency channels to subtract out the small foreground contamination
in the central channels.  In Planck, the difference between the predicted
and actual rms noise from falsely assuming perfect frequency coherence
can be more than two orders of magnitude.

For Boomerang and Planck, the largest residual noise component,
aside from detector noise, is dust emission and is sufficiently
large that one might worry that current uncertainties in our knowledge
of the foreground model may affect
the implications for the detection of the SZ effect.  It is therefore
important to explore variations on our fiducial foreground 
model. 

Multiplying
the foreground rms amplitudes uniformly by a factor of 2 (and hence
the power by a factor of 4), produces less than a factor of 2 increase
in the residual noise rms as shown in Fig.~\ref{fig:simul}. 
Likewise, as discussed in \cite{Tegetal99} (1999), minor
variations in the frequency coherence do not effect the residual noise much
in spite of the fact that it is crucial not to assume perfect
correlation. 
We conclude that uncertainties in the properties of currently known
foregrounds are unlikely to change our conclusions qualitatively.
There is however always the possibility that some foreground that does not
appear in the currently-measured frequency bands will affect our
results.

The fact that the residual dust contributions are comparable to those 
of the detector noise for Boomerang and Planck is problematic for
another reason.  Since the algorithm minimizes
to total residual variance, it attempts to keep these two main
contributors roughly comparable.  However the dust will clearly 
be non-Gaussian to some extent and one may prefer instead to trade
more residual detector noise for dust contamination.  One can
modify the subtraction algorithm to account for this by artificially
increasing the rms amplitude of the dust when calculating the weights in
equation~(\ref{eqn:weights}) while using the real amplitude 
in calculating the residual noise in equation~(\ref{eqn:residualcomponent}).
For example we have explored increasing the amplitude by a factor of
4 (power by 16) for the weights.  The result is an almost negligible increase 
in total residual noise rms but an improvement in dust rejection by 
a factor of 3-4 in rms.   For Planck this brings the ratio of 
dust to total rms to $\sim 10\%$ and recall that the noise adds
in quadrature so that the total dust contribution is really 
$\sim 1\%$ of the total.
This more conservative approach is thus advisable but since
it leaves the total residual noise rms essentially unchanged, we
will adopt the minimum variance noise to estimate the
detection threshold.

\begin{figure}[b]
\centerline{\psfig{file=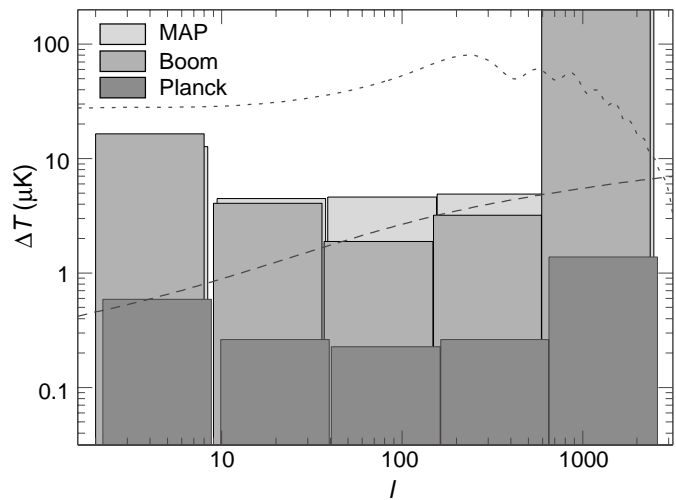,width=3.5in,angle=0}}
\caption{Detection thresholds for the SZ effect.   Error boxes represent the
1-$\sigma$ rms residual noise in multipole bands and can be interpreted 
as the detection threshold.  Also shown (dotted) is the level of the primary
anisotropies that have been subtracted with the technique and the signal
(dashed) expected in the simplified model of \S \ref{sec:sz}.}
\label{fig:error}
\end{figure}

Fig.~\ref{fig:clean} directly tells us the detection threshold per
$(l,m)$ multipole moment.  Since the SZ signal is likely to have a smooth 
power spectrum in $l$,
one can average over bands in $l$ to beat down the residual noise.  
Assuming Gaussian-statistics, the residual noise variance $ 2N_l^2$ for
the power spectrum estimate is then given by
\begin{equation}
N_l^{-2}\Big|_{\rm band} = 
{f_\sky}  \sum_{l_{\rm band}} (2l+1) N_l^{-2}\,,
\end{equation}
where $f_\sky$ accounts for the reduction of the number of independent 
modes due to the fraction of sky covered.
The result for the three experiments is shown in Fig.~\ref{fig:error}.
In the absence of a detection, they can be interpreted as the optimal
1 $\sigma$ upper limits on SZ bandpowers achievable by the experiment.
Boomerang and MAP can place upper limits on the SZ signal in the interesting
$\mu$K regime whereas Planck can detect signals well below a $\mu$K.

This noise averaging procedure in principle implicitly assumes that the 
statistical properties of the residual noise, and by implication the
full covariance matrix of the other foregrounds, is precisely 
known.  In reality, they too must be estimated from the multifrequency data 
itself through
either through the subtraction techniques discussed here or
by direct modeling of the foregrounds in the maps.  \cite{Tegetal99} (1999)
found that direct modeling of the foregrounds with hundreds of fitted
parameters did not appreciably degrade our ability to extract the
properties of the primary anisotropies.  The main source of variance there
was the cosmic variance of the primary anisotropies themselves whose 
properties are precisely known.
Similarly here the main source of residual variance is either the 
primary anisotropies (for MAP) or detector noise (for Boomerang and Planck)
and their statistical properties may safely be considered known.

\section{Modeling the SZ Signal}
\label{sec:sz}

In order to estimate how well the statistical properties of 
the SZ effect might be recovered with multifrequency CMB maps, 
we need to model the large-angle SZ effect itself. 
The current state-of-the-art in hydrodynamic simulations
(\cite{daS99} 1999; \cite{Refetal99} 1999; \cite{Seletal00} 2000)
has reached a qualitative but not quantitative consensus on the
statistical properties of the SZ effect.   In addition,
questions as to the heating of the gas from non-gravitational
sources may even change the results qualitatively (\cite{Pen99} 1999).
Hydrodynamic simulations are also severely limited in the 
dynamic range and volume sampled.  

Given the current
state of affairs, we believe it is useful to explore a 
parameterized model of the effect whose consequences are
simple to calculate and which may be calibrated against hydrodynamic
simulations as they continue to improve.

\subsection{Bias Prescription}

In general, the SZ temperature fluctuation $\Theta=\Delta T/T$ 
is given by
the opacity weighted
integrated pressure fluctuation along 
the line of sight:
\begin{eqnarray}
\Theta^\sz(\bn,\nu) 
	       &=& -2 s(\nu) \int_0^{\rad_0} d\rad\
			\dot\tau \pi(r,\bn r) \,,
\end{eqnarray}
$\rad$ is the the comoving distance, $\tau$ is the Thomson optical depth, 
overdots are derivatives with respect to $\rad$ and
the dimensionless electron pressure fluctuation is
\begin{equation}
\pi = \delta p_e/\rho_e\,.
\end{equation}
One needs to model the statistical properties of $\pi$, in particular
its power spectrum and bispectrum
\begin{eqnarray}
\left< \pi(\bfk)^* \pi(\bfk') \right> &=&  (2\pi)^3 \deld(\bfk -\bfk')P_\pi(k) 
		\,,  \\
\left< \pi(\bfk)\pi(\bfk')\pi(\bfk'') \right> &=& (2\pi)^3 \deld(\bfk+\bfk'+\bfk'')
		B_\pi(k,k',k'')\,,\nonumber
\end{eqnarray}
as a function of lookback time or distance $r$.   In principle we also need the
unequal time correlators, but in practice these do not play a role 
as we shall see.

By analogy to the familiar case of galaxy clustering, 
it is reasonable to 
suppose that the pressure fluctuations depend locally
on the dark matter density and hence are biased
tracers of the dark matter density in the {\it linear} regime (\cite{GolSpe99} 1999).
Hence the statistical properties follow from those of the dark matter distribution
\begin{eqnarray}
P_\pi(k;r) &\approx& b_\pi(r)^2 P_\delta(k;r)\,,    \nonumber\\
B_\pi(k,k',k'';r) &\approx& b_\pi(r)^3 B_\delta(k,k',k'';r)   \,. 
\end{eqnarray}
We have restored the time dependence since the bias will be time dependent
even in the linear regime and must be extracted from simulations. 
In general, the bias parameter for the power spectrum and bispectrum
need not be the same even in the linear regime since the bispectrum
automatically involves higher order corrections (\cite{FryGaz93} 1993).  
For estimation purposes here we will take them to be equal. 

Following \cite{GolSpe99} (1999), we chose the form
\begin{eqnarray}
b_\pi(r) = b_\pi(0) /(1+z) \, ,
\end{eqnarray}
as motivated by findings that the average gas temperature 
drops off roughly by this factor.  We normalize the value of 
the bias parameter today by comparison with recent
hydrodynamic simulations.  It is conceptually useful to separate
the bias into two factors: 
\begin{equation}
b_\pi(0) = {k_B T_e(0) \over m_e c^2} b_{\delta} \,,
\end{equation}
i.e. an opacity-weighted average temperature and a bias parameter
for the gas density at that temperature.
In \cite{Refetal99} (1999), for our fiducial $\Lambda$CDM cosmology,
the bias $b_\delta$ was found to be 
$\sim$ 8 to 9, while in \cite{Seletal00} (2000) it was
found to be in the range $\sim$ of $3$ to $4$.  
In both these papers, $T_e(0) \sim$ 0.3 to 0.4 keV; 
these values are lower than the $\sim$ 1 keV found by
\cite{CenOst99} (1999) using hydrodynamical simulations with
feedback effects. 
As a compromise between these results, we take  
$T_e(0)=0.5$keV and $b_\delta=4$, which corresponds to   
\begin{equation}
b_\pi(0) = 0.0039\,.
\end{equation}
Note that this is a factor of 2 lower than used in
\cite{GolSpe99} (1999) and \cite{CooHu99} (1999).

Needless to say, the resulting predictions should be taken
as order-of-magnitude estimates only.
As simulations 
improve, one can expect better values for the bias today and a more detailed 
modeling of its redshift and perhaps even scale dependence.

\subsection{Multipole Moments}

The multipole moments of the SZ effect under these simplifying assumptions
can then be expressed 
as a weighted projection of the
density field (\cite{CooHu99} 1999):
\begin{eqnarray}
\almn^\sz(0) &\equiv&
	\int d \bn Y_l^{m*}(\bn)\Theta^\sz(\bn,0)\nonumber\\
     &\approx& i^l \int {d^3{\bf k } \over 2\pi^2} \delta({\bf k},r_l) I_l^\sz(k) 
     \Ylmn{}^*(\bk)  \, ,
\label{eqn:szsource}
\end{eqnarray}
where 
\begin{eqnarray}
I_l^\sz(k) &\approx& W^{\sz}(\rad_l) \sqrt{{\pi \over 2l}}{1
\over k} F_l(k) \, , \nonumber\\
W^\sz(r) & = & -2 b_\pi(r) \dot\tau\,,
\end{eqnarray}
in the Limber approximation
and
(\cite{Hu00a} 2000a)
\begin{eqnarray}
\rad_l   & = & \Omega_K^{-1/2} H_0^{-1} \sinh^{-1} (\Omega_K^{1/2} H_0 l/k) \,,\nonumber\\
F_l   & = & (1 +\Omega_K H_0^2 l^2/k^2)^{-1/4}  \,.
\label{eqn:Limber}
\end{eqnarray}
The quantities take on a simple forms for a flat universe: $\rad_l \rightarrow l/k$ and
$F_l(k) \rightarrow 1$. 
The Limber approximation breaks down for $l \la 50$ but is sufficient for our purposes. 

\begin{figure}[t]
\centerline{\psfig{file=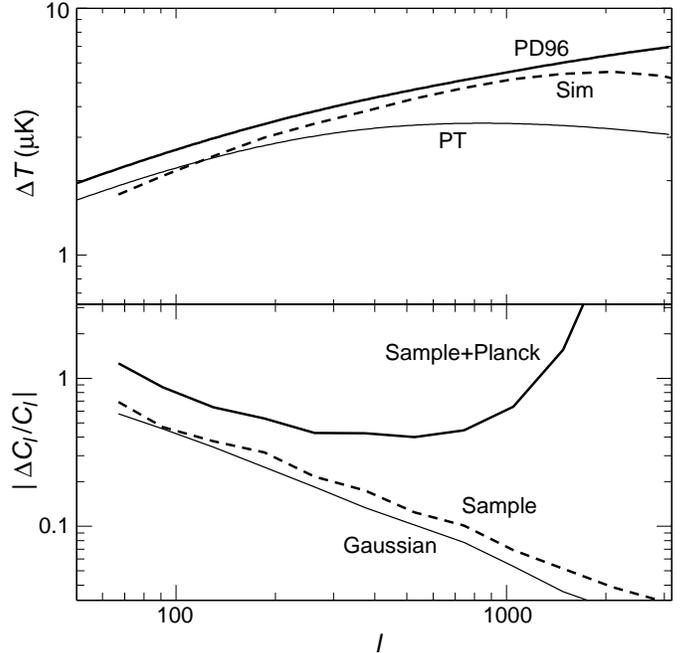,width=3.5in,angle=0}}
\caption{{\it top:} SZ power spectrum from simulations
compared to analytical predictions based on linear perturbation 
theory (PT) and the non-linear scaling relations of \protect\cite{PeaDod96} (1996; PD96). {\it bottom:} 
Errors on the binned power spectrum estimators for a single 
$6^\circ\times 6^\circ$ field; for a given experiment
the errors should be scaled by $\sim 0.03 f_\sky^{-1/2}$.  
The sampling errors in the simulations is nearly equal to
those of a Gaussian random field with the same power spectrum.
The total noise including residual foregrounds and detector noise
is also given for Planck.  }
\label{fig:simultwopt}
\end{figure}

\subsection{Power Spectrum}

The power spectrum of the SZ effect in this simplified model follows
from equation~(\ref{eqn:szsource}),
\begin{eqnarray}
C_l^{\sz} &=& {2 \over \pi} \int {dk \over k} {k^3
P_\delta\left(k;r_l\right)} [I_l^\sz(k)]^2 
		\nonumber\,,\\
	  &\approx& \int_0^{\rad_0}d\rad {[W^\sz(r)]^2  \over \da^2 }
		P_\delta(l/\da;r) \,,
\label{eqn:clsz}
\end{eqnarray}
In the second line we have transformed the integration variable 
under the Limber correspondence: $k=l/\da$ and 
\begin{equation}
\int {d k\over k} F_l^2\ldots  = \int {d\rad \over \da}\ldots \; .
\end{equation}
We see that to go from the flat to curved cosmologies in the Limber
approximation one simply replaces the radial distance with the angular
diameter distance in the integrand. 

In evaluating the SZ power spectrum, we have extended the SZ model
to the non-linear regime 
by using the scaling formulae for the nonlinear dark matter power
spectrum of  \cite{PeaDod96} (1996).
However, modeling the SZ effect with a scale-independent bias factor will clearly break
down deep in the non-linear regime. \cite{Refetal99} (1999) have shown that it is 
a reasonable approximation in the weakly non-linear regime (overdensities $\la 10$)
for $z \la 1$ but can be in serious error outside of this range. As 
the weakly non-linear regime is the one of interest for 
anisotropies at $l \la 1000$, we will use this
approximation to test the effects of non-linearities.
The predicted power spectrum in our fiducial model is shown 
in Fig.~\ref{fig:simultwopt}.

\subsection{Bispectrum}
\label{sec:bispectrum}

The bispectrum of the SZ effect also follows from
expression~(\ref{eqn:szsource}) 
\begin{eqnarray}
\bilm &\equiv& \left<a_{l_1m_1} a_{l_2m_2} a_{l_3m_3}\right>  \nonumber\\
&=&
\left[ \prod_{j=1}^3 i^{l_j} \int \frac{d^3 k_j}{2\pi^2} 
I_{l_j}^\sz(k_j)  Y_l^{m\ast}(\hat{\bf k}_j)  \right]
\nonumber \\
&& 
(2\pi)^3 \deld(\veck_1+\veck_2+\veck_3)
B_\delta(k_1,k_2,k_3) \,.
\nonumber
\end{eqnarray}
Here the density bispectrum should be understood as arising from the full 
unequal time correlator
\begin{equation}
\left< \delta(\bfk_1;r_1) \delta(\bfk_2;r_2) \delta(\bfk_3;r_3) \right> \,,
\end{equation}  
where the temporal coordinate, which we temporarily suppress,
 is evaluated in the Limber approximation
(\ref{eqn:Limber}). 

To further simplify this expression, we expand the delta function 
\begin{equation}
\deld(\veck_1+\veck_2+\veck_3) = \frac{1}{(2 \pi)^3} \int e^{i
(\veck_1+\veck_2+\veck_3) \cdot  \bn r} d^3x \, ,
\end{equation}
and employ the Rayleigh expansion
\begin{equation}
e^{i \veck \cdot \bn r} = 4 \pi \sum_{lm} i^l j_l(k r) Y_l^{m
\ast}(\hat{\veck}) Y_l^m(\bn) \, .
\end{equation}
We have assumed here a flat universe to simplify the derivation; as we
have seen in the last section, we can promote the final result to a 
curved universe by replacing radial distances with angular diameter 
distances.

With these relations, the angular integral over the directions of
$\bfk_j$ collapse to give
\begin{eqnarray}
\bilm
&=& \int r^2 dr \left[ \prod_{j=1}^3 {2 \over \pi} \int k_j^2 dk_j
I_{l_j}^\sz(k_j) j_{l_j}(k_j r) \right]   \nonumber\\
&& \times B(k_1,k_2,k_3) G_{l_1 l_2 l_3}^{m_1 m_2 m_3}\,,
\end{eqnarray}
where the Gaunt integral is
\begin{eqnarray}
G_{l_1 l_2 l_3}^{m_1 m_2 m_3} &\equiv& \int d\bn
      \Ylm{1} \Ylm{2} \Ylm{3} \\
&=&\sqrt{(2l_1+1)(2l_2+1)(2l_3+1)  \over 4\pi}
\nonumber\\ &&\times
      \wj \wjm \,. \nonumber 
\label{eqn:harmonicsproduct}
\end{eqnarray}
Here, the quantities in parentheses are the Wigner-3$j$ symbols
whose properties are described in Appendix A of \cite{CooHu99} (1999).
The integrals over the Bessel functions can again be done in 
the Limber approximation 
leaving
\begin{eqnarray}
\bilm
&=& G_{l_1 l_2 l_3}^{m_1 m_2 m_3} \int dr 
	{[W^\sz(r)]^3 \over r^4}  
B_\delta({l_1 \over r},{l_2 \over r},{l_3\over r};r)\,, \nonumber
\end{eqnarray}
Note that only equal time contributions contribute in the Limber approximation.

We can promote this result to a curved universe by 
replacing radial distances with angular diameter distances
\begin{eqnarray}
\bilm 
&=& G_{l_1 l_2 l_3}^{m_1 m_2 m_3} 
	\int dr {[W^\sz(r)]^3 \over \da^4} 
		B_\delta({l_1 \over \da},{l_2 \over \da},{l_3\over \da};r) \,.
\nonumber
\end{eqnarray}
Finally, we can introduce the angular averaged bispectrum as
\begin{eqnarray}
\bi = \sum_{m_1 m_2 m_3}  \wjm \bilm\,,
\end{eqnarray}
to obtain the final result
\begin{eqnarray}
\bi &=& \sqrt{(2l_1+1)(2l_2+1)(2l_3+1)  \over 4\pi} \wj
	\nonumber\\
&&\times \int dr {[W^\sz(r)]^3 \over \da^4}  B_\delta({l_1 \over \da},{l_2 \over \da},{l_3\over \da};r) \,.
\label{eqn:szbispectrum}
\end{eqnarray}	
One can alternately derive this relation by taking a flat-sky 
approach and using the general relation between the flat-sky and
all-sky bispectra (see Appendix C, \cite{Hu00b} 2000b).

Equation (\ref{eqn:szbispectrum}) gives the SZ angular bispectrum in
terms of the underlying density bispectrum.  In second order
perturbation theory, the density bispectrum is in turn given by
\begin{eqnarray}
B_\delta (k_1,k_2,k_3;r) &=& F_2(\bfk_1,\bfk_2) P_\delta (k_1;r) P_\delta (k_2;r)  \nonumber\\
&& + 5\; {\rm perm.}\,,
\end{eqnarray}
where
\begin{eqnarray}
F_2(\bfk_1,\bfk_2) = \frac{5}{7} + \frac{\veck_1 \cdot \veck_2}{k_2^2}
+ \frac{2}{7} \frac{(\veck_1 \cdot \veck_2)^2}{k_1^2\ k_2^2}\,.
\end{eqnarray}

Unfortunately, there exists no accurate fitting formula for the bispectrum of the
density field in the mildly non-linear regime; we will employ simulations in \S \ref{sec:sims}
to address this regime.  In the deeply non-linear regime,
the density field obeys the hierarchical ansatz
\begin{eqnarray}
B_\delta (k_1,k_2,k_3;r) = {Q_3 \over 2} [P(k_1;r)P(k_2;r)+{\rm 5\;  perm.}]\,,
\end{eqnarray}
where the power 
spectra are given by the non-linear scaling of \cite{PeaDod96} (1996).
\cite{ScoFri99} (1999) find that for power law power spectrum
\begin{eqnarray}
Q_3 (n) = [4 - 2^n]/[1+2^{n+1}] \,.  
\end{eqnarray}
\cite{Hui99} (1999) suggests that for a general power spectrum
one should replace $n$ with the local linear power spectral index at $(k_1+k_2+k_3)/3$.

\begin{figure}[b]
\centerline{\psfig{file=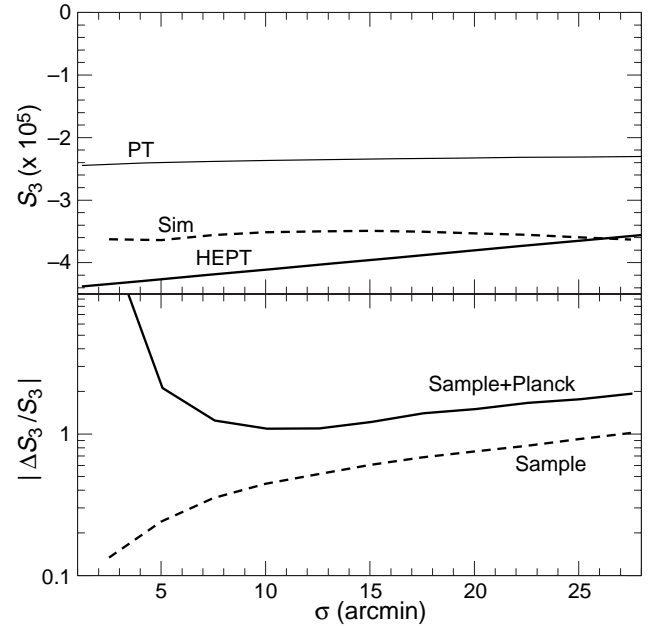,width=3.5in,angle=0}}
\caption{{\it top:} Skewness in the simulations 
compared with second order perturbation theory
(PT)  and hyper-extended perturbation theory (HEPT). 
The smoothing is performed with an 
angular
tophat of radius $\sigma$.
{\it bottom:}
Errors on the skewness measurement for a single
$6^\circ\times 6^\circ$ field due to sampling errors
and residual noise from Planck.}
\label{fig:skewness}
\end{figure}

\subsection{Skewness}

The simplest aspect of the bispectrum that can be measured is the
third moment of the map smoothed on some scale with a window $W(\sigma)$
\begin{eqnarray}
\left< \Theta^3(\bn;\sigma) \right> &=&   
		{1 \over 4\pi} \sum_{l_1 l_2 l_3}
		\sqrt{(2l_1+1)(2l_2+1)(2l_3+1) \over 4\pi} \nonumber\\
		&&\times \wj \bi W_{l_1}(\sigma)W_{l_2}(\sigma)W_{l_3}(\sigma)
		\,,
\end{eqnarray}
where $W_l$ are the multipole moments (or Fourier transform in a flat-sky approximation)
of the window.   For simplicity, we will choose windows which are either
top hats in real or multipole space.

It is useful to define the skewness parameter
\begin{equation}
S_3(\sigma) =  {\left< \Theta^3(\bn;\sigma) \right> \over \left< \Theta^2(\bn;\sigma) \right>^2}\,,
\end{equation}
where 
the second moment is that of the SZ signal
\begin{equation}
\left< \Theta^2(\bn;\sigma) \right> =  
{1 \over 4\pi} \sum_l (2l+1) C_l^\sz W_l^2(\sigma)\,.
\end{equation}
The skewness in our fiducial model is shown 
for both the perturbation theory and HEPT predictions in
Fig.~\ref{fig:skewness}.

Since the {\it density} bispectrum in both the perturbative and non-linear
regime scale as $[P_\delta(k)]^2$, the amplitude of the underlying density 
fluctuations roughly scale out of $S_3$.  However, the pressure bias $b_\pi$
does {\it not}: $S_3 \propto b_\pi^{-1}$.  
$S_3$ thus provides an observable handle on the bias.
This general point applies even if the bias is non-linear although its
interpretation will be not be as straightforward (see \cite{FryGaz93} 1993 and
\cite{MoJinWhi97} 1997 for its application in galaxy biasing).   

\begin{figure}[b]
\centerline{\psfig{file=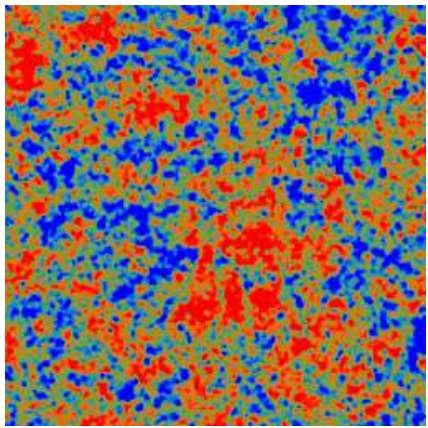,width=2.2in,angle=0}}
\caption{One of 500 simulations 
of the SZ effect in the $\Lambda$CDM model for a 
$6^\circ\times 6^\circ$ 
field-of-view.  The range of the map is $-100\mu K , 25\mu K$ with an
rms of $9\mu K$ and has an approximate angular resolution of $2'$.
Note the lack of obvious filamentary structures.}
\label{fig:simulmap}
\end{figure}

\begin{figure*}
\centerline{
\psfig{file=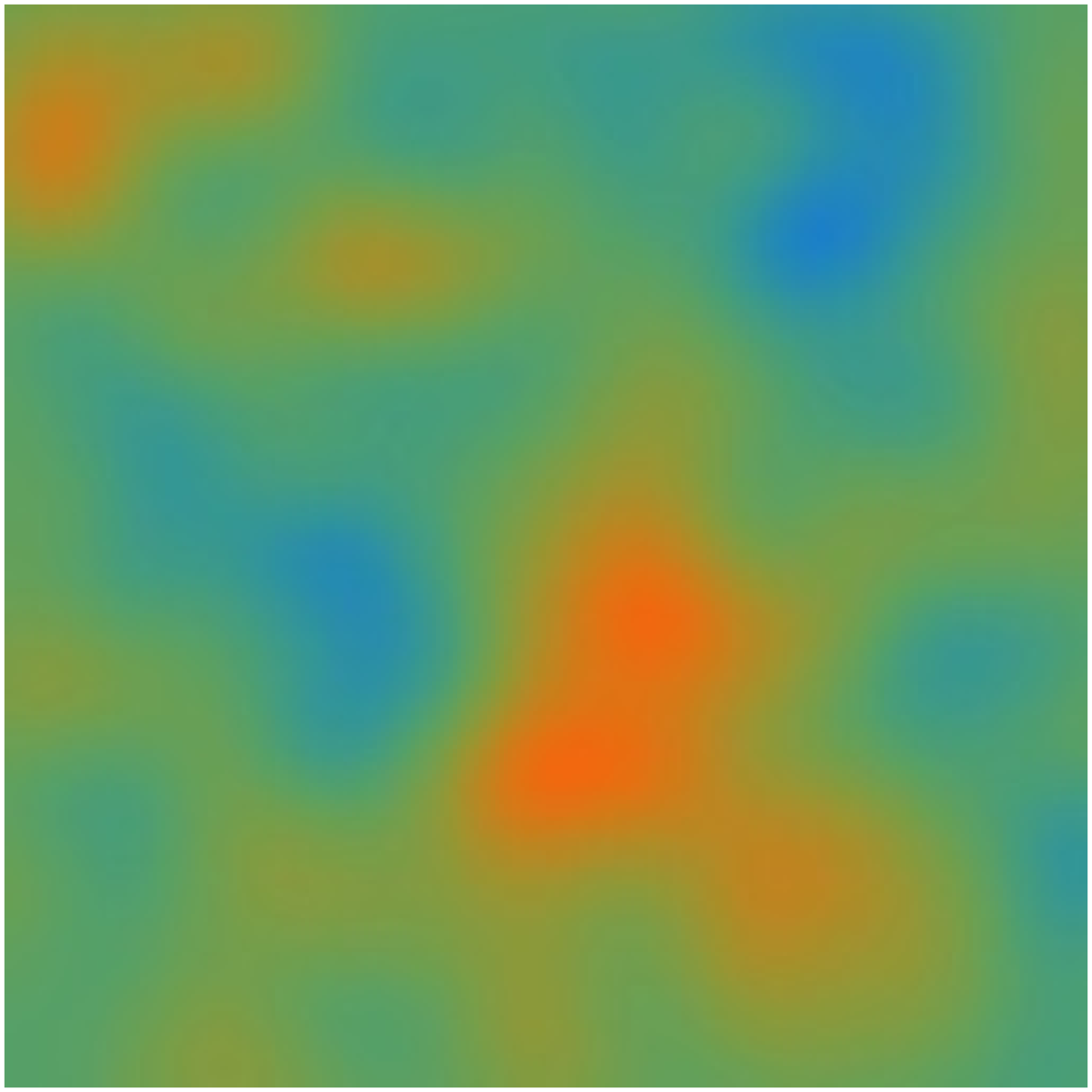,width=2.2in,angle=0}
\psfig{file=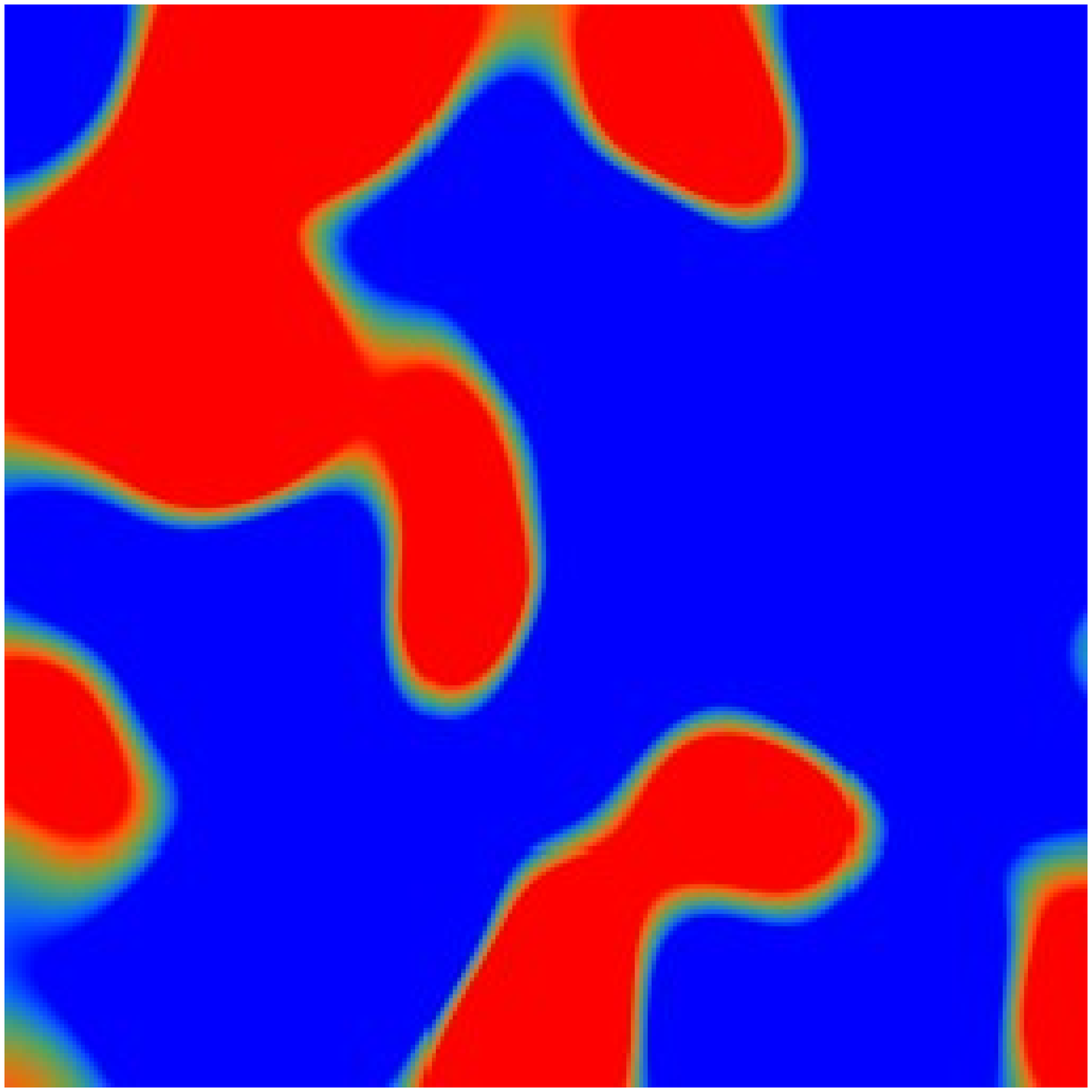,width=2.2in,angle=0}
\psfig{file=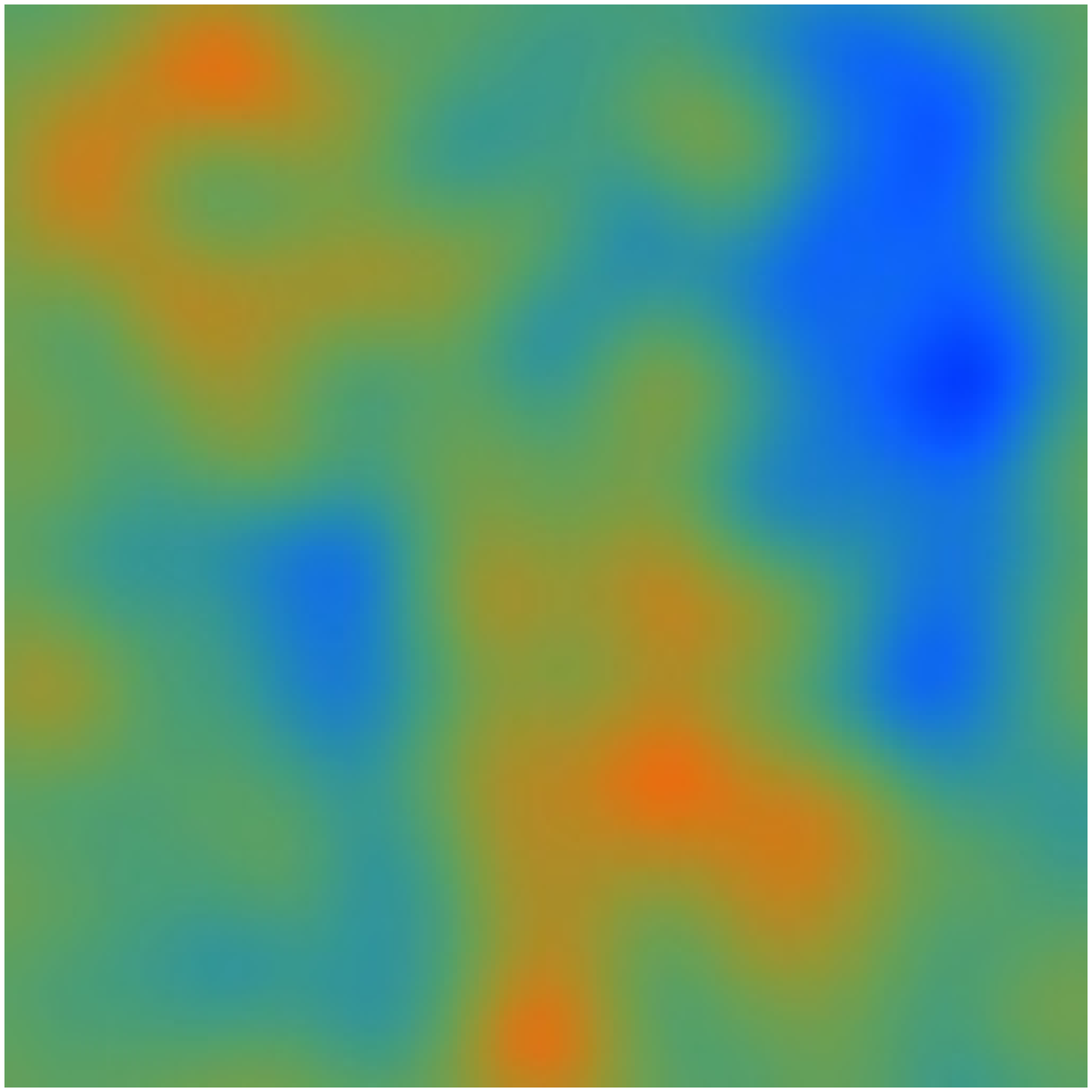,width=2.2in,angle=0}
	}
\caption{Recovery of the SZ signal with Planck: left to right, model
SZ signal, signal $+$ noise from primary anisotropies and foregrounds, and
final recovered map from Planck.   The signal map is that
of Fig.~\protect\ref{fig:simulmap} smoothed with a top-hot of
radius $20'$.
}
\label{fig:recovery}
\end{figure*}

\subsection{Numerical Simulations}
\label{sec:sims}

Since we are interested in the properties of the SZ effect in the 
weakly-nonlinear regime, 
cosmological simulations are required to recover the complete statistical
properties of the signal and calibrate
semi-analytic approaches for its low-order statistics.  
The simplified SZ model employed in this paper 
has the virtue that it is easy to simulate as it requires only dark matter 
and not the gas to model.  Its main drawback of course is that results must
be taken with a grain of salt due to missing physics.

The realism of the basic
approach can be improved by better calibrating the bias model 
against hydrodynamic simulations.  One can envision 
going beyond the simple redshift dependent bias approach taken 
here to include scale dependence and stochasticity.   Even accounting
for these additional complications, simple dark matter simulations can
continue to complement full hydrodynamic simulations.   
Hydrodynamic simulations will always
be more limited in dynamic range and sampling volume.  
Indeed, the current state of the art is limited
a handful of realizations across one order of magnitude in physical scale 
(\cite{Refetal99} 1999; \cite{Seletal00} 2000).  A single simulation
is then ``stacked'' on the line-of-sight.
Given the range of redshifts at which the SZ effect contributes, the
simulation volume is traced many times for each line-of-sight.  Moreover,
the angular resolution decreases monotonically as one approaches the
origin at $z=0$. 

The reduction in dynamic range due to the angular projection is a
serious but not unfamiliar problem in cosmology.  It occurs whenever
the kernel for the projection spans cosmological distances. 
\cite{WhiHu99} (1999) introduced a technique of tiling multiple particle-mesh simulations which
telescope along the line of sight to maintain a fixed angular resolution for
the analogous problem in weak lensing.   
This also avoids the problem of over-representing the filamentary structure of
the map noted by \cite{Refetal99} (1999).  

We refer the reader to 
\cite{WhiHu99} (1999) for details of the approach and tests of the method.
The simulation all have a 
$256^3$ mesh with $256^2$ lines of sight for the ray tracing on a $6^\circ \times 6^\circ$ field.  
Other relevant parameters are given in Tab.~\ref{tab:sims}:
the box size $L_{\rm box}$, the number of particles $N_{\rm part}$, the number of simulations run
$N_{\rm sim}$, the number of tiles of the given box size used $N_{\rm tile}$, the maximum
redshift to which a given box is used, and the angular resolution of the mesh the maximum and
minimum redshift used $\theta_{\rm mesh}$.  
Note that we cannot shrink the box size along the line-of-sight indefinitely since
the fundamental mode of the box must be in the linear regime to provide accurate evolution.
This implies that we lose angular resolution near the origin where a fixed physical scale
subtends a large angle on the sky. 
Furthermore at the higher redshift the number of particles must be increased to eliminate
shot noise from the initial conditions.  Nonetheless, the tiling technique does a good
job of maintaining angular resolution at all but the lowest redshifts.

We construct 500 SZ maps from random combinations of the tiles in Tab.~1 for our 
statistical analysis;  one realization is shown in Fig.~\ref{fig:simulmap}.
The average power spectrum is shown in
Fig.~\ref{fig:simultwopt} (top panel) and compared with the linear perturbation
theory prediction and the non-linear scaling relation of \cite{PeaDod96} (1996).
We have tested that the deficit of power at the low
multipoles is an artifact of the finite field-of-view through monte-carlo
realizations of the predicted power spectrum.   The roll-off at high
multipoles is due to the spatial resolution in the simulations.  This 
also explains the $\sim 10\%$ deficit at intermediate scales which 
comes from highly non-linear structure close to the origin.   Agreement
is restored if one eliminates contributions from overdensities $>10$ in
the predictions.  Since our SZ model is at best valid in the weakly 
non-linear regime, these contributions should not be included anyway. 

Fig.~\ref{fig:skewness} (top panel) shows the results for the skewness
in the simulations compared with the second order perturbation theory
and HEPT predictions.  The agreement here is worse,  but is still
sufficient for our purposes, given the crudeness of the underlying
model for the SZ effect itself. 

We can address sample variance questions from the scatter of the
results in the individual realizations.   Sampling errors for
the power spectrum and skewness are shown in the bottom panels of
Fig.~\ref{fig:simultwopt} and \ref{fig:skewness} respectively.
Since these are for individual $6^\circ \times 6^\circ$ planes,
they should be scaled by $\sim 0.03 f_\sky^{-1/2}$ for a given 
experiment.  Sampling errors are one source of noise that
we will include in the signal-to-noise calculations in the
next section.

\begin{table}[tb]\footnotesize
\caption{\label{tab:sim}}
\begin{center}
{\sc Details of Numerical Simulations}
\begin{tabular}{cccccc}
\tableskip\hline\tableskip
$L_{\rm box}$ & $N_{\rm part}$ & $N_{\rm sim}$ & $N_{\rm tile}$ & $z_{\rm max}$ & $\theta_{\rm mesh}$ \\
\tableskip\hline
445 &$256^3$ & 5 & 2 & 3.00  & $1.4'-1.8'$    \\
355 &$256^3$ & 5 & 2 & 1.87  & $1.4'-1.8'$    \\
280 &$256^3$ & 5 & 2 & 1.27  & $1.4'-1.8'$    \\
220 &$256^3$ & 5 & 2 & 0.90  & $1.4'-1.8'$    \\
175 &$128^3$ & 6 & 2 & 0.66  & $1.4'-1.8'$    \\
140 &$128^3$ & 6 & 2 & 0.50  & $1.4'-1.8'$    \\
110 &$128^3$ & 6 & 2 & 0.38  & $1.4'-1.8'$    \\
85  &$128^3$ & 6 & 2 & 0.29  & $1.4'-1.8'$    \\
70  &$128^3$& 10 & 9& 0.22  & $1.4'-\infty$ \\
\tableskip\hline
\end{tabular}
\end{center}
NOTES.---%
Numerical simulations in our $\Lambda$CDM cosmological model; see
text for a description of these quantities.
\label{tab:sims}
\end{table}

\begin{figure}[t]
\centerline{\psfig{file=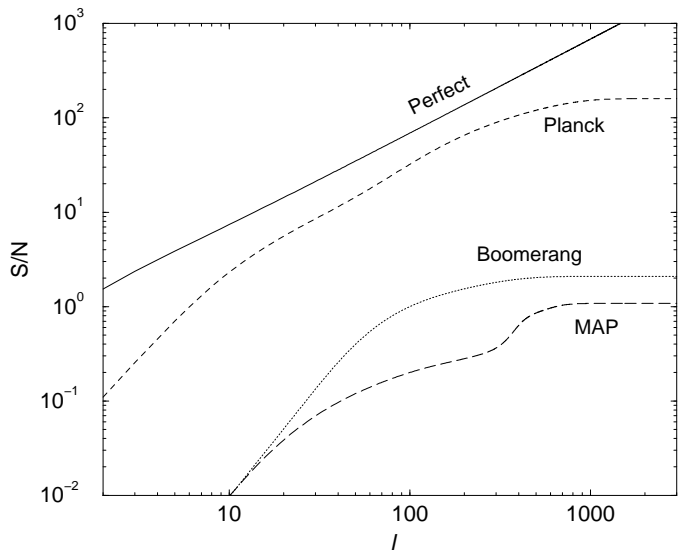,width=3.5in,angle=-90}}
\caption{Cumulative signal-to-noise in the measurement of the  SZ
power spectrum with Boomerang, MAP and Planck as a  function of maximum
$l$. The solid line
is the maximum signal-to-noise achievable in a perfect experiment
(see text).
}
\label{fig:clsn}
\end{figure}

\section{Estimating the Signal-to-Noise}
\label{sec:sn}
With the SZ signal estimated from the simple bias model of \S \ref{sec:sz}
and residual noise calculated from the foreground model and subtraction
techniques of \S \ref{sec:cleaning}, we can now estimate the signal-to-noise
for the detection of the SZ effect.  
In Fig.~\ref{fig:recovery}, we illustrate the foreground subtraction
technique on simulated Planck maps.  The signal-to-noise in the maps
is of order one for features spanning tens of arcminutes.  We shall here
show that this level of signal-to-noise is more than sufficient for
the purpose of extracting measurements of the low order statistics
of the SZ signal. 

\subsection{Power Spectrum}

The signal-to-noise in the power spectrum per multipole $(l,m)$ mode is simply
\begin{equation}
\left( {S \over N} \right)_{l m}^2  = {1 \over 2} 
	\left( {C_l^\sz \over {C_l^\tot}}\right)^2 \, .
\end{equation}
Here, $C_l^\tot$ is the  power spectrum of all contributions in 
the SZ map,
\begin{eqnarray}
C_l^{\rm tot} = C_l^\sz + N_l\,,
\label{eqn:cltot}
\end{eqnarray}
where recall that the residual noise $N_l$ 
was defined in equation (\ref{eqn:nl}) and includes contributions
from detector noise.

Assuming Gaussian statistics for the signal and noise, each mode
is independent so that the total signal-to-noise is the quadrature
sum 
\begin{equation}
\left( {S \over N} \right)^2 = {f_\sky \over 2}\sum_l (2l+1) \left( 
	{C_l^\sz \over {C_l^\tot}} \right)^2 \, .
\end{equation}
This quantity gives the variance of the total power measurement in the
SZ effect, including sample variance.  ${\rm S/N} \gg 1$ means that 
one has a precise measurement of the power spectrum not simply
a highly significant detection.
In Fig.~\ref{fig:clsn}, for the Boomerang, MAP
and Planck experiments as a function of the maximum $l$ mode included
in the sum. We also show the ultimate limit of perfect foreground and
noise removal where $C_l^{\rm tot}=C_l^\sz$ and $f_\sky=1$. 
We will refer to this case here and below as a ``perfect experiment''.

With our fiducial
choice of the gas bias, Planck should have a highly significant
detection of the total signal.  One should
bear in mind that the bias parameter $b_\pi$ is still highly
uncertain and that the $S/N$ scales as $b_\pi^2$.  Nevertheless even a 
relatively large reduction in the average gas temperature
or density bias will not make the signal undetectable in principle. 
In practice, however remember that one is then relying on a precise 
subtraction of the noise bias in the measurement of $C_l^\tot$, 
which in turn requires that the power spectrum of the dust and other 
residual foregrounds lurking at least at the $10\%$ 
level in rms  (1\% in power)
are determined comparably precisely.   

If the fiducial SZ bias is close to correct, the high total single-to-noise 
in Planck can be used to break
the measurement into
bands in $l$ and recover the band power with errors 
\begin{equation}
\left( \Delta C_l^\sz \over C_l^\sz \right)^{-2} =  {f_\sky \over 2}
\sum_{l_{\rm band}} (2l+1)
\left( { C_l^\sz \over C_l^\tot}\right)^2 \,.
\end{equation}
We give an example from monte carlo realizations of the Gaussian noise
and sample variance from the simulations in Fig.~\ref{fig:simultwopt}.
Note that these are errors for a $6^\circ \times 6^\circ$ section
of the sky and should be scaled by $\sim 0.03 f_\sky^{-1/2} \approx
0.04$ for Planck.

These signal-to-noise estimates assume that both the signal and noise
are Gaussian.  Of course in reality the SZ signal is non-Gaussian.
In general, gravitational collapse correlates the amount of power 
in density fluctuations across all scales in the non-linear regime.
However since the SZ effect probes many 
independent density fluctuations along the line-of-sight, the central
limit theorem ensures that the SZ signal is far more Gaussian than
the density field. We can test how much this affects the signal-to-noise 
with our simulations. Shown in Fig.~\ref{fig:simultwopt} is the 
sampling errors on the band powers from the simulations themselves 
as compared with those from Gaussian realizations of the same power spectrum.  
The excess variance over the Gaussian limit is small on the relevant scales 
given detector
noise limitations from Planck.   

\begin{figure}[t]
\centerline{\psfig{file=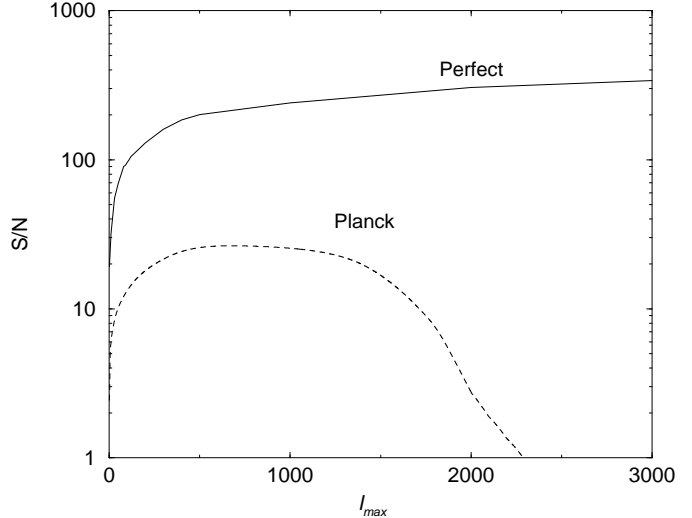,width=3.5in,angle=-90}}
\caption{Cumulative signal-to-noise in the measurement of the third
moment, $\Theta^3$, with top hat smoothing in multipole space (i.e. truncation
above $l_{\rm max}$).  The HEPT approximation to the bispectrum
is assumed here. MAP and Boomerang (not shown) have signal-to-noise
values less than $0.1$ everywhere.}
\label{fig:t3}
\end{figure}

\subsection{Skewness}
\label{sec:skewness}

The overall signal-to-noise for the measurement of the third moment of
SZ effect is
\begin{equation}
\left( {S \over N} \right)^2 = {f_\sky}
		{ \left< \Theta^3(\bn;\sigma) \right>^2 \over {\rm Var}}
\end{equation}
where the variance is given by 
\begin{eqnarray}
{\rm  Var} &=& { 1 \over (4\pi)^2 } \sum_{l_1 l_2 l_3}
		{(2l_1+1)(2l_2+1)(2l_3+1) \over 4\pi} \wj^2   \nonumber\\
		&& \times W_{l_1}^2(\sigma) W_{l_2}^2(\sigma) W_{l_3}^2(\sigma)
				6 C_{l_1}^\tot C_{l_2}^\tot C_{l_3}^\tot\,.
\label{eqn:t3var}
\end{eqnarray}
In Fig.~\ref{fig:t3}, we show the signal-to-noise for a measurement of
the third moment as calculated under the HEPT. We compare the
signal-to-noise in Planck with the ideal case of perfect
removal of foregrounds and detector noise, and full sky coverage.  We use here a tophat window
in multipole space out to $l_{\rm max}$ 
to conform with other signal-to-noise considerations.
Cosmic variance and Planck detector
noise reduces the signal-to-noise values both at the
low and high end for $l_{\rm max}$ values respectively. For Planck, the $l$ values in the
range of few hundred to $\sim$ 1000 provides the maximal signal-to-noise for a
measurement of the skewness.  This corresponds to smoothing scales $\sigma 
\sim$ 10'-30' for tophat windows in angular space 
(c.f.~Fig.~\ref{fig:skewness}). For MAP and Boomerang, the
signal-to-noise values are $\lesssim$ 0.1, suggesting that a
detection of SZ skewness is not likely to be possible in these two experiments.

Again equation~(\ref{eqn:t3var}) assumes Gaussian statistics for
the variance and ignores the sample variance of the third moment 
itself.  We test this approximation in Fig.~\ref{fig:skewness} and
find that it is reasonable given the level of residual noise for Planck. 
In constructing an estimator for $S_3$, it is important
to remove the noise bias since noise variance will always reduce the
skewness in the map.  We do this by multiplying the estimator by
$(\left<\Theta_{\rm tot}^2 \right>/\left<\Theta_{\rm SZ}^2 \right>)^2$.

Finally, note that in the noise-dominated regime
the signal-to-noise in $S_3$ scales strongly with the gas bias
$S/N \propto b_\pi^3$, so that the detectability of this effect
depends strongly on currently uncertain assumptions.

\begin{figure}[t]
\centerline{\psfig{file=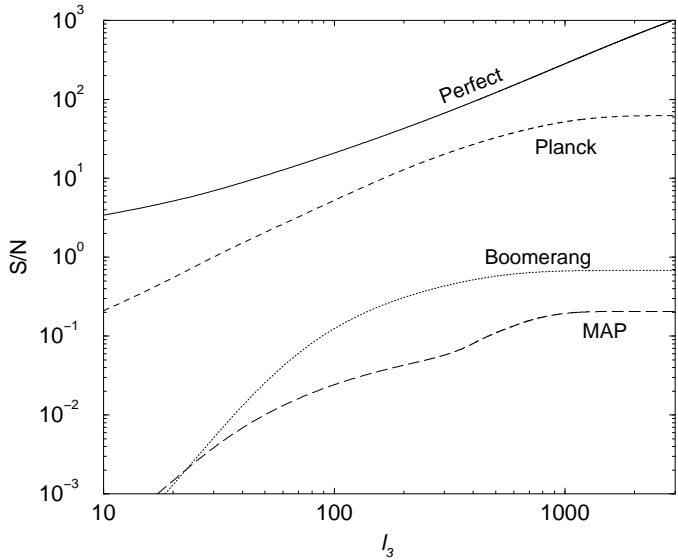,width=3.5in,angle=-90}}
\caption{Cumulative signal-to-noise for the detection of 
SZ bispectrum as a function of $l_3$ multipole. The solid line
is the maximum signal-to-noise achievable  in a perfect experiment.
}
\label{fig:szbispec}
\end{figure}

\subsection{Bispectrum}

The full bispectrum of the SZ effect contains all of the information
about its three-point correlations induced by the growth of structure
beyond the linear approximation.   The skewness is simply one,
easily measured, aspect of the bispectrum.
The full signal-to-noise ratio of the bispectrum is
\begin{equation}
\left( {S \over N} \right)^2 = 
f_\sky \sum_{l_1,l_2,l_3} 
	{\bi^2 \over 
		 6 C_{l_1}^\tot   C_{l_2}^\tot   C_{l_3}^\tot   } \, ,
\label{eqn:bispecnoise}
\end{equation}
where $C_l^\tot$ follows Eq.~(\ref{eqn:cltot}). We plot the bispectrum
cumulative signal-to-noise  as a function of signal $l_3$, summed over
$l_1$ and $l_2$.
We refer the reader to \cite{CooHu99} (1999) for a detailed discussion on the 
bispectrum, its variance and the calculation of signal-to-noise ratio.

In Fig.~\ref{fig:szbispec}, we show the expected cumulative 
signal-to-noise for the SZ bispectrum in Boomerang, MAP and Planck data and
a perfect experiment.
The signal-to-noise is
calculated
under the HEPT approximation for the underlying density field. 
As shown, MAP and Boomerang
allow reasonable limits to be placed on any non-Gaussian signal in the SZ
effect while Planck allows a strong possibility for a detection.

Again the same caveats as to the sensitivity of the $S/N$ estimate
to the underlying assumptions
that applied for the skewness also apply here.  Moreover, measuring
all the configurations of the bispectrum will be a formidable 
computational challenge as will control over systematic effects in the
experiments.

\subsection{Lensing Correlation}

The SZ effect and weak
gravitational lensing of the CMB both trace large-scale structure in
the underlying density field.  
By measuring the correlation, one can directly test the
manner in which gas pressure fluctuations trace the dark matter
density fluctuations.  
The correlation vanishes in the two-point functions since the
lensing does not affect an isotropic CMB due to conservation
of surface brightness. 

The correlation manifests itself as a 
non-vanishing bispectrum in the CMB at RJ
frequencies (\cite{GolSpe99} 1999; \cite{CooHu99} 1999).   
Again
the cosmic variance from the primary anisotropies presents an 
obstacle for detection of the effect above the several arcminute
scale ($l\sim 2000$).
With the multifrequency cleaning of the
SZ map presented here one can enhance the detectability of the
effect. 

Consider the bispectrum composed of one $a_{l m}$ from the cleaned
SZ map and the other two from the CMB maps.  Call this the SZ-CMB-CMB
bispectrum.   The noise variance of this term will be reduced
by a factor of $C_l^\tot / C_l^\cmb$ compared with the
CMB-CMB-CMB bispectrum.  As one can see from Fig.~\ref{fig:clean}
this can be up to a factor of $10^3$ in the variance.
Details for the calculation of the CMB-CMB-CMB bispectrum
are given in \cite{CooHu99} (1999).  Here we 
have updated the normalization for SZ effect, taken
$f_{\sky}=0.65$ for Planck's useful sky coverage, and
compared the $S/N$ of the two bispectra.
As shown, the measurement using foreground cleaned Planck SZ and CMB
maps has a substantially higher signal-to-noise than  that from
using the Planck CMB map alone for multipoles $l \sim 10^2-10^3$.

Beyond the improvement in signal-to-noise, however, 
there is an important 
advantage in constructing the SZ-lensing bispectrum using SZ and CMB
maps. A mere measurement of the bispectrum in CMB data can lead to
simultaneous detection of non-Gaussianities through processes other than just
SZ-lensing cross-correlation. As discussed in \cite{GolSpe99} (1999)
and extended in \cite{CooHu99} (1999), gravitational lensing
also correlates with other late time secondary anisotropy contributors
such as integrated Sachs-Wolfe (ISW; \cite{SacWol67} 1967) effect
and the reionized Doppler effect. In addition to lensing correlations,
non-Gaussianities can also  be generated through reionization and
non-linear growth of perturbations (\cite{SpeGol99} 1999;
\cite{GolSpe99} 1999; \cite{CooHu99} 1999). 
Bispectrum measurements at a signle frequency can result in a confusion as to the relative contribution from
each of these scenarios. In \cite{CooHu99} (1999), we
suggested the possibility of using differences in individual bispectra as
a function of multipoles, 
however, such a separation can be problematic 
given that these differences are subtle (e.g.,
Fig~6 of \cite{CooHu99} 1999). 

The construction
of the SZ-lensing bispectrum using SZ and CMB maps has the advantage
that one eliminates all possibilities, other than SZ, that result in a
bispectrum. For effects related to SZ,
the cross-correlation of lensing and  SZ should produce the dominant
signal; as shown in \cite{CooHu99} (1999), bispectra signal through SZ
and reionization effects, such as Ostriker-Vishniac (OV;
\cite{OstVis86} 1986), are considerably smaller.

Conversely, multifrequency cleaning also eliminates the SZ 
contribution from the CMB maps and hence a main contaminant of 
the CMB-CMB-CMB bispectrum.
This assists in the detection of smaller signals such as
the 
ISW-lensing correlation, Doppler-lensing correlation or the 
non-Gaussianity of the initial conditions. 
Such an approach is
highly desirable and Planck will allow such detailed studies to be
carried out.

A potential caveat is that as noted above, the full bispectrum 
in an all-sky satellite experiment will be difficult to measure.
\cite{ZalSel99} (1999) have developed a reduced set of three-point statistics
optimized for lensing studies, based on a two point reconstruction of
the lensing-convergence maps from temperature gradient information.  
They show that most of the information
is retained in these statistics.  Multifrequency cleaning improves
the signal-to-noise for these statistics by exactly the same factor
as for the full bispectrum.

\begin{figure}
\centerline{\psfig{file=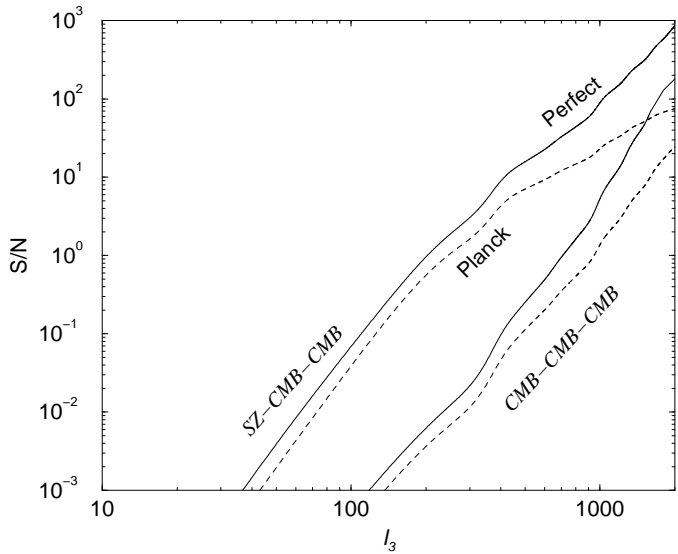,width=3.5in,angle=-90}}
\caption{Cumulative signal-to-noise in the measurement of the  SZ-weak
gravitational lensing cross-correlation through the bispectrum
measurement in CMB data.  Compared are the
expected signal-to-noise with (SZ-CMB-CMB) and without (CMB-CMB-CMB) 
multifrequency isolation
of the SZ effect for Planck and a perfect/cosmic variance limited
experiment.   
Multifrequency isolation provides additional
signal-to-noise and the opportunity to uniquely identify the bispectrum
contribution with the SZ effect.}
\label{fig:szlens}
\end{figure}

\section{Discussion}
\label{sec:discussion}

We have studied the prospects for extracting the statistical properties
of the Sunyaev-Zel'dovich (SZ) effect associated with hot gas in 
large-scale structure using upcoming multifrequency CMB experiments.  
This gas currently remains undetected but may comprise a substantial
fraction of the present day baryons. 
The SZ effect has a distinct spectral dependence with a null at a frequency of
$\sim$ 217 GHz compared with true temperature anisotropies. 
This frequency dependence is what allows for effective separation
of the SZ contribution with multifrequency
CMB measurements.  

As examples, we have employed the frequency and noise specifications  
of the Boomerang, MAP, Planck experiments. 
The MAP satellite only covers frequencies at 
RJ part of the frequency spectrum.  Consequently, only
Boomerang and Planck can take full advantage of multifrequency separation of
the SZ and primary anisotropies.  We have evaluated the
detection threshold for SZ power
spectrum measurements (see Fig.~\ref{fig:error}).
Boomerang and MAP should provide limits on the degree
scale fluctuations at the several   
$\mu$K level in rms; Planck should be able to detect sub $\mu$K signals.

The expected level of the SZ signal in the fiducial $\Lambda$CDM
model is still somewhat uncertain.  We have employed a simple
bias model for the pressure fluctuations, roughly normalized
to recent hydrodynamic simulations
(\cite{Refetal99} 1999; \cite{Seletal00} 2000), and calculated
the resulting signal using analytic scaling relations and particle-mesh 
dark matter simulations.
As hydrodynamic simulations improve, these techniques 
can be extended with more sophisticated modeling of the
bias.  They complement hydrodynamic simulations by
extending the dynamic range and simulated volume, the latter
being important for questions of sample variance. 

Assuming this simplified model of the SZ signal, Planck should
have signal-to-noise per multipole of order unity
for $l < 1000$.  Although the recovered maps are then somewhat 
noisy, they are sufficient for precise determinations of 
low order statistics such as the SZ power spectrum, bispectrum
and skewness (see Figs.~\ref{fig:simultwopt}-\ref{fig:szbispec}).
The skewness in principle can be used to separate the pressure bias
from the underlying amplitude of the density fluctuations.
The full bispectrum contains significantly more information but will
be difficult to extract in its entirety.
Current methods for measuring the bispectrum, tested with the 
COBE data, have concentrated at measuring specific modes such as
$l_1 = l_2 =l_3 = l$ (\cite{Feretal98} 1998).  
More work will clearly be required, 
especially in understanding the systematic errors at a 
sufficient level, but the wealth of information potentially
present in the bispectrum should motivate efforts.

Note however that the non-Gaussianity in the SZ signal is not very strong
due to the fact that it is constructed from many independent pressure
fluctuations along the line of sight.  As a consequence, we expect that
signal-to-noise ratios can be estimated by Gaussian approximations, 
but that techniques that try to improve the SZ-primary separation
based on non-Gaussianity (\cite{Hobetal98} 1998; \cite{AghFor99} 1999)
may not be particularly effective for this signal.  

We caution the reader that our oversimplification of the SZ signal
can cause problems for a naive interpretation of future detections.
For example, \cite{Seletal00} (2000) find that the SZ power spectrum
in their simulations is dominated by shot noise from the rare hot 
clusters not included in our modeling.
Fortunately since these contributions are highly non-Gaussian, they can 
can readily be identified and removed.   At the very least, $X$-ray
bright clusters can be externally identified and removed; this has
been shown to substantially reduce the shot noise contribution (\cite{KomKit99} 1999).
The effect we are modeling should be understood as the signal
in fields without such clusters.  

Another means of separating the SZ signal from large-scale structure
from that of massive clusters is to cross correlate it with other 
tracers of large-scale structure that are less sensitive to highly
overdense regions.   
An added benefit is that such a cross-correlation will 
also empirically measure the extent to which pressure fluctuations
follow mass fluctuations.
The CMB anisotropies themselves
carry one such tracer in the form of the convergence from 
weak lensing.  It manifests itself as a three-point correlation 
or bispectrum (\cite{GolSpe99} 1999) but without frequency
information it is severely sample-variance limited due to 
confusion noise from primary anisotropies.  
Measuring the SZ-lensing correlation using the cleaned SZ maps
improves the signal-to-noise for the detection by over an order
of magnitude at degree scales.   Furthermore, the techniques
introduced by \cite{ZalSel99} (1999) provide a concrete algorithm
for extracting most of the three-point signal without
recourse to measuring all the configurations of the bispectrum. 
Conversely, SZ removal from
the CMB maps themselves can assist in the detection of
other smaller bispectrum signals by eliminating one source
of confusion noise.

The cross-correlation coefficient between the SZ effect and CMB weak
lensing is relatively 
modest ($\sim$ 0.5, see \cite{Seletal00} 2000). This is due to
the fact that the SZ effect is a tracer of the nearby universe while CMB
lensing is maximally sensitive to structure at $z\sim 3$. A higher
correlation is expected if SZ is cross-correlated with an
external probe of low redshift structure.
\cite{PeiSpe00} (2000)
suggested the cross-correlation of MAP CMB data and
Sloan\footnote{http://www.sdss.org} galaxy data.
An improved approach would be to use the Planck derived SZ map
rather than a CMB map.
Using a SZ map reduces noise from the primary anisotropies
and guarantees that any
detection is due to correlations with the SZ effect. 
Extending the calculations in
\cite{PeiSpe00} (2000) with the Planck generated
SZ map, we
find signal-to-noise ratios which are on average greater by a factor of
$\sim$ 10 when compared to signal-to-noise values using MAP CMB map.
In fact with redshifts for galaxies, Planck SZ map can be
cross-correlated in redshifts bins to study the 
redshift evolution of the gas. 
Other promising possibilities include cross correlation with 
soft X-ray background measurements,
as well as ultraviolet and soft X-ray absorption line studies.

All these considerations imply a bright future for
SZ studies of the hot gas associated with large-scale structure
with wide-field multifrequency CMB observations. 
Its detailed properties should be revealed in its non-Gaussianity and
correlation with other tracers of large-scale structure.

\acknowledgments
We thank Martin White for permission to adapt his PM ray tracing 
code for these purposes.  We acknowledge 
useful discussions with  Lloyd Knox, Joe Mohr, Roman Scoccimarro, Ned Wright and
Matias Zaldarriaga. ARC is grateful to John 
Carlstrom, Michael Turner and Don York for helpful advice and financial
support. WH is supported by the Keck Foundation, a Sloan Fellowship,
and NSF-9513835. MT acknowledges NASA grant NAG5-6034 and Hubble
Fellowship HF-01084.01-96A from STScI, operate by AURA,  Inc. under
NASA contract NAS5-26555. We acknowledge the use of CMBFAST
(\cite{SelZal96} 1996).

\end{document}